\documentclass[sigconf]{acmart}
\AtBeginDocument{%
  }

\setcopyright{acmlicensed}
\copyrightyear{2026}
\acmYear{2026}
\setcopyright{cc}
\setcctype{by}
\acmConference[KDD 2026] {Proceedings of the 32nd ACM SIGKDD Conference on Knowledge Discovery and Data Mining V.1}{August 9--13, 2026}{Jeju Island, Republic of Korea.}
\acmBooktitle{Proceedings of the 32nd ACM SIGKDD Conference on Knowledge Discovery and Data Mining V.1 (KDD 2026), August 9--13, 2026, Jeju Island, Republic of Korea}
\acmISBN{979-8-4007-2258-5/2026/08}
\acmDOI{10.1145/3770854.3785688} 
\usepackage{booktabs} 
\usepackage{amsmath}
\usepackage{amsfonts}
\usepackage{comment}
\usepackage{footmisc}
\usepackage{mathrsfs}
\usepackage{graphicx}
\usepackage{subfigure}
\usepackage{multirow}
\usepackage{algorithm}
\usepackage{algorithmic}
\usepackage{multicol}
\usepackage{enumitem}
\usepackage{array}
\usepackage{float}
\usepackage{hyperref}
\usepackage{xspace}
\usepackage{xcolor}
\usepackage[table]{xcolor}
\definecolor{lightgreen}{RGB}{220,255,220}
\definecolor{lightred}{RGB}{255,220,220}
\definecolor{lightorange}{RGB}{255,235,205}
\usepackage{pifont}
\newcolumntype{P}[1]{>{\raggedright\arraybackslash}p{#1}}
\usepackage{tcolorbox}  
\tcbuselibrary{breakable}   
\usepackage{balance}

\newcommand{\name}{A/B Agent\xspace}
\newcommand{\AB}{A/B testing\xspace}

\newcommand{\etc}{\emph{etc.}\xspace}

\begin{document}

\title{Exploring Recommender System Evaluation: A Multi-Modal User Agent Framework for A/B Testing}

\author{Wenlin Zhang}
\authornote{Contributed equally to this work.}
\orcid{0000-0003-1809-8264}
\affiliation{%
  \institution{City University of Hong Kong}
  \city{Hong Kong}
  \country{China}
}
\email{wl.z@my.cityu.edu.hk}

\author{Xiangyang Li}
\authornotemark[1]
\orcid{0000-0003-2862-0239}
\affiliation{%
  \institution{Huawei Technologies Ltd.}
  \city{Shenzhen}
  \country{China}
}
\email{lixiangyang34@huawei.com}

\author{Qiyuan Ge}
\orcid{0009-0008-0569-1855}
\affiliation{%
  \institution{City University of Hong Kong}
  \city{Hong Kong}
  \country{China}
}
\email{qiyuange2-c@my.cityu.edu.hk}

\author{Kuicai Dong}
\orcid{0000-0002-5564-0641}
\affiliation{%
  \institution{Huawei Technologies Ltd.}
  \city{Shenzhen}
  \country{China}
}
\email{dong.kuicai@huawei.com}

\author{Pengyue Jia}
\orcid{0000-0003-4712-3676}
\affiliation{%
  \institution{City University of Hong Kong}
  \city{Hong Kong}
  \country{China}
}
\email{jia.pengyue@my.cityu.edu.hk}

\author{Xiaopeng Li}
\orcid{0009-0008-6162-8500}
\affiliation{%
  \institution{City University of Hong Kong}
  \city{Hong Kong}
  \country{China}
}
\email{xiaopli2-c@my.cityu.edu.hk}

\author{Zijian Zhang}
\orcid{0000-0003-1194-8334}
\affiliation{%
  \institution{Jilin University}
  \city{Changchun}
  \country{China}
}
\email{zhangzijian@jlu.edu.cn}

\author{Maolin Wang}
\orcid{0000-0002-0073-0172}
\affiliation{%
  \institution{City University of Hong Kong}
  \city{Hong Kong}
  \country{China}
}
\email{morin.wang@my.cityu.edu.hk}

\author{Yichao Wang}
\orcid{0000-0001-7053-8269}
\authornote{Corresponding authors.}
\affiliation{%
  \institution{Huawei Technologies Ltd.}
  \city{Shenzhen}
  \country{China}
}
\email{wangyichao5@huawei.com}

\author{Huifeng Guo}
\orcid{0000-0002-7393-8994}
\affiliation{%
  \institution{Huawei Technologies Ltd.}
  \city{Shenzhen}
  \country{China}
}
\email{huifeng.guo@huawei.com}

\author{Ruiming Tang}
\orcid{0000-0002-9224-2431}
\affiliation{%
  \institution{Huawei Technologies Ltd.}
  \city{Shenzhen}
  \country{China}
}
\email{tangruiming@huawei.com}

\author{Xiangyu Zhao}
\authornotemark[2]
\orcid{0000-0003-2926-4416}
\affiliation{%
  \institution{City University of Hong Kong}
  \city{Hong Kong}
  \country{China}
}
\email{xianzhao@cityu.edu.hk}

\renewcommand{\shortauthors}{Wenlin Zhang et al.}
\begin{abstract}
  In recommender systems, online \AB is a crucial method for evaluating the performance of different models. However, conducting online \AB often presents significant challenges, including substantial economic costs, user experience degradation, and considerable time requirements. With the Large Language Models' powerful capacity, LLM-based agent shows great potential to replace traditional online \AB. Nonetheless, current agents fail to simulate the perception process and interaction patterns, due to the lack of real environments and visual perception capability. To address these challenges, we introduce a multi-modal user agent for \AB (\name). Specifically, we construct a recommendation sandbox environment for \AB, enabling multimodal and multi-page interactions that align with real user behavior on online platforms. The designed agent leverages multimodal information perception, fine-grained user preferences, and integrates profiles, action memory retrieval, and a fatigue system to simulate complex human decision-making. We validated the potential of the agent as an alternative to traditional \AB from three perspectives: model, data, and features. Furthermore, we found that the data generated by \name can effectively enhance the capabilities of recommendation models.
 Our code is publicly available at \url{https://github.com/Applied-Machine-Learning-Lab/ABAgent}.
\end{abstract}

\ccsdesc[500]{Information systems~Recommender systems}

\keywords{Recommender System, Multimodal User Agent, A/B Testing}

\maketitle

\section{Introduction}
\label{sec:introduction}

\begin{figure}
	\centering
	\includegraphics[width=1.0\linewidth]{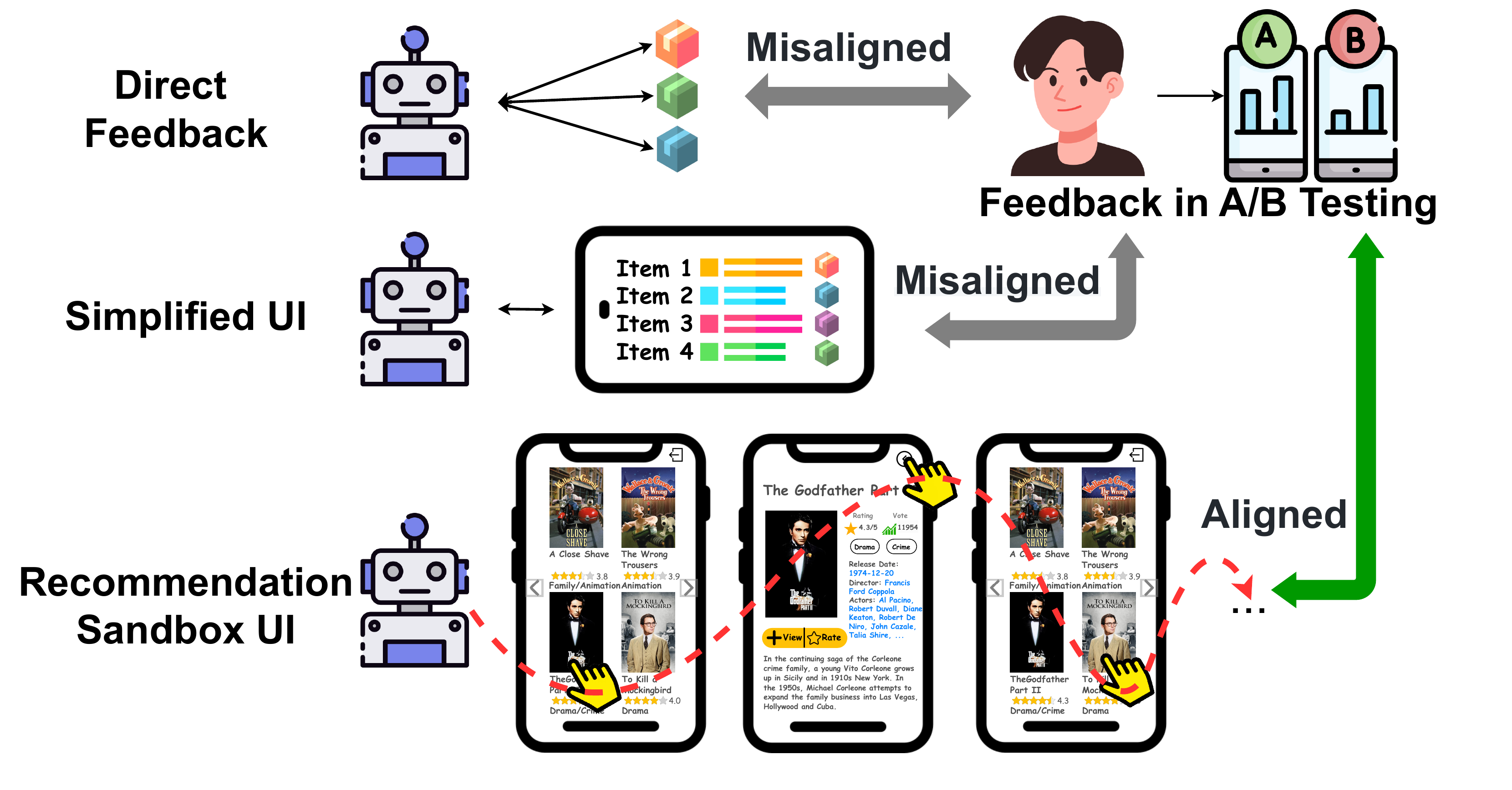}
	\caption{Comparison of alignment between simulated user feedback and real user feedback in \AB.}
	\label{fig:Fig1_Motivation}
\end{figure}

In real-world industrial settings, recommender systems, spanning diverse paradigms from reinforcement learning~\cite{zhao2018deep, zhao2018recommendations, liu2023exploration, zhao2020jointly} and sequential modeling~\cite{liu2023linrec, li2023strec, gao2024smlp4rec} to emerging large model-based approaches~\cite{fu2025unified, zhang2025notellm}, often rely on online \AB to optimize and evaluate model performance in real time~\cite{kohavi2015online, xu2015infrastructure, nandy2021b, fabijan2018online}.
In \AB pipeline, users are randomly divided into experimental and control groups, with the experimental group receiving recommendations from a new system, while the control group uses the existing system as a baseline. However, despite its advantages, A/B testing has several challenges: 
1) High resource consumption: It requires significant server resources and data analysis efforts, especially for high-traffic products~\cite{gilotte2018offline}. 2) User experience degradation: Frequent A/B testing can disrupt user interaction and lower satisfaction~\cite{li2012unbiased}. 3) Evaluation latency 
: Collecting sufficient data for statistical significance can delay timely evaluation.  
Therefore, it is essential to develop reliable offline evaluation methods to simulate \AB results. 

However, replicating the complex dynamics of A/B testing poses significant challenges for user simulator design. In real-world settings, users interact with multimodal content across various interfaces, actively explore items of interest, disengage when fatigued, and generate feedback across multiple interaction stages. Capturing these behaviors accurately is crucial for building effective user simulators that reflect real-world evaluation dynamics.

With the emergence of Large Language Models (LLMs), LLM-based agents have garnered significant attention. 
LLM agents have broad world knowledge, require less scenario-specific training, and enable flexible, interpretable interactions through natural language~\cite{wang2024survey, lin2023agentsims, zhu2023ghost, wang2023voyager, zhang2025personalize}.
For instance, iEvaLM~\cite{wang2023rethinking} employs an LLM-based user agent to evaluate conversational recommender systems, providing flexible natural language interactions. RecAgent~\cite{wang2023user} designs a Web recommendation scenario where user agents can interact with each other, exploring the impact of human social behavior on recommendation results. Agent4Rec~\cite{zhang2024generative} designed a user agent for simulating page-by-page movie recommendations.
These works demonstrate that LLM-based agents have the promising capability to simulate user behaviors.

Although existing work has achieved active interaction between agents and recommendation environments, there remains a significant gap between simulated paths and real human perception process. Figure~\ref{fig:Fig1_Motivation} illustrates the simulation methods used in existing approaches. They either directly simulate user-item interaction feedback or model user behavior in simplified, text-based UI environments. Apparently, they fail to accurately reflect real user behavior on recommendation platforms. Moreover, the agents lack perception of rich multimodal information~\cite{dong2025mmdocir} and multi-layered interface simulation, which limits their ability to replicate human interactions. To address this gap, we propose \name to simulate human perception process and interaction path more effectively.


To design \name, two challenges need to be tackled. \textbf{(1) The gap between the simulated environment and the online platform UI.} Existing simulated environments directly present only textual item information to users, overlooking the fact that users in an actual online platform UI obtain multimodal information at different granularities and explore progressively. To simulate the process of how users perceive and interact with the UI on an online platform, we crawled multimodal movie data and constructed a movie recommendation sandbox environment similar to IMDB.
\textbf{(2) Modeling user behavior trajectories under multimodal and multi-level information settings.} 
In the sandbox environment, users interact with movie information at varying levels of granularity across different interfaces, which results in complex behavior trajectories. However, existing user agent designs struggle to achieve fine-grained perception and human-like exploration. To address these challenges, we developed \name, an agent that simulates human perception and exploration more effectively. For perception, \name captures detailed user preferences across diverse levels of movie information granularity, integrating both image and text modalities. For exploration, \name incorporates a long-term and short-term memory module and a fatigue system to avoid repeated exploration or overexploitation.

Finally, to verify the effectiveness of the agent's simulation in the sandbox recommendation environment, we conducted \AB using the agent for different recommendation algorithms. 
Furthermore, we collected feedback data from the agent during the interaction process for data augmentation experiments to validate the impact of simulated data on model improvement. Experimental results demonstrate that \name can emulate user interaction patterns in the interactive recommendation environment.
Our key contributions can be summarized as follows:
\begin{itemize}[left=0pt, itemindent=0pt, itemsep=0pt, topsep=0pt]
    \item  We propose \name to simulate the entire human perception process and interactive behavior trajectories for \AB. The agent possesses multimodal information perception and can perform human-like exploration within the sandbox recommendation environment.
    \item We develop an interactive sandbox recommendation environment, where the agent retrieves movie information at varying levels of granularity across different interfaces, enabling it to perform multi-interface exploration. 
    \item  We create a large-scale multimodal dataset MM-ML-1M by extending movies' meta-information and posters. This dataset can provide the necessary data for different interfaces within the interactive sandbox recommendation environment. 
    \item We conduct extensive experiments to evaluate the agent's simulation capabilities within the sandbox environment. Both \AB for recommendation models and data augmentation based on agent feedback demonstrate the effectiveness of \name.
\end{itemize}


\begin{figure*}
	\centering
	\includegraphics[width=1.0\linewidth]{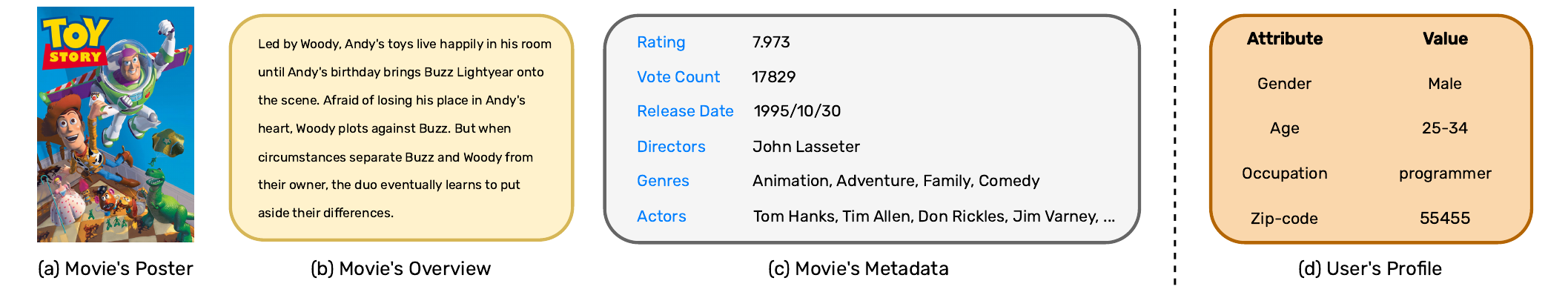}
	\caption{Data example from Multimodal-Movielens-1M dataset.
    }
	\label{fig:Fig2_Dataset}
\end{figure*}
\section{\name Framework}
\subsection{Framework Overview}
\label{framework_overview}
As shown in Figure~\ref{fig:Fig2_Agent} 
the \name framework is composed of three primary components:
1) \textbf{MM-ML-1M Dataset}: This comprehensive dataset includes images, text, and various movie metadata, providing a realistic basis for simulating authentic movie recommendation scenarios. 2) \textbf{Recommendation Sandbox Environment}: This environment offers a wide range of popular recommendation models and a rich user interface, enabling interaction and exploration by the User Agent in a simulated real-world setting. 3) \textbf{\name}: Designed to emulate user behavior patterns in a realistic movie recommendation environment, it consists of several key components, including the agent profile module, memory module, action module, and fatigue system. These systems enable the agent to deliver feedback akin to human responses.

\subsection{Dataset: MM-ML-1M}
Most existing recommendation datasets fall short in effectively simulating real-world interaction scenarios. For instance, Amazon lacks user attribute information, MovieLens~\cite{harper2015movielens} is devoid of multimodal data, and many datasets like Criteo~\cite{zhu2021open} and Avazu~\cite{zhu2021open} have anonymized feature characteristics. 
To address these limitations, we introduce the Multimodal-MovieLens-1M (MM-ML-1M) dataset, an extension of the original MovieLens-1M~\cite{harper2015movielens}, enhanced with additional movie posters and metadata. Figure~\ref{fig:Fig2_Dataset} provides a detailed illustration of the data structure.
Table~\ref{tab:Tab1_dataset} summarizes the statistics of the MM-ML-1M dataset, including the type, count, and value range for each feature. For text fields, the range denotes the character length. The dataset contains 6,040 users and 3,952 movies, with an interaction sparsity of 4.19\%.
\begin{table}[t]
\caption{Dataset statistics in MM-ML-1M. Note: the range for text type is the length of text content.
}
\label{tab:Tab1_dataset}
\resizebox{\columnwidth}{!}{%
\begin{tabular}{cccc}
\toprule
\textbf{Feature} & \textbf{Type} & \textbf{Count} & \multicolumn{1}{c}{\textbf{Range}} \\ 
\midrule
Title            & Text          & 3,822              & {[}2, 72{]}                        \\
Overview         & Text          & 3,814              & {[}13, 991{]}                      \\
Genres           & Text          & 3,789              & {[}5, 64{]}                        \\
Rating           & Numerical     & 3,822              & {[}0, 10{]}                    \\
Vote Count       & Numerical     & 3,822              & {[}0, 30,002{]}                     \\
Release Date     & Date          & 3,820              & {[}1911-05-05, 2024-06-07{]}       \\
Directors        & Text          & 3,810              & {[}3, 172{]}                       \\
Actors           & Text          & 3,785              & {[}8, 5,101{]}                      \\
Poster           & Image         & 3,814              & - 
                                   \\ 
\midrule
\textbf{User Count}   & - & 6040     &  - \\
\textbf{Movie Count}  & - & 3952     &  - \\
\textbf{Sparsity}     & - & 0.0419   &  - \\
\bottomrule
\end{tabular}
}
\end{table}
MM-ML-1M enriches the movie-side information with elements such as posters, overviews, and metadata, including IMDb ratings, vote counts, directors, and actors, while maintaining the original user-side information. These enhancements offer crucial visual~\cite{dong2025mmdocrag} and contextual data, improving recommendation simulations by capturing factors like movie popularity and creator preferences. This comprehensive dataset facilitates the development and evaluation of recommendation models that more accurately reflect real-world user interactions.

\subsection{Recommendation Sandbox Environment}
\label{section_sandbox}
\begin{figure*}
	\centering
	\includegraphics[width=0.95\linewidth]{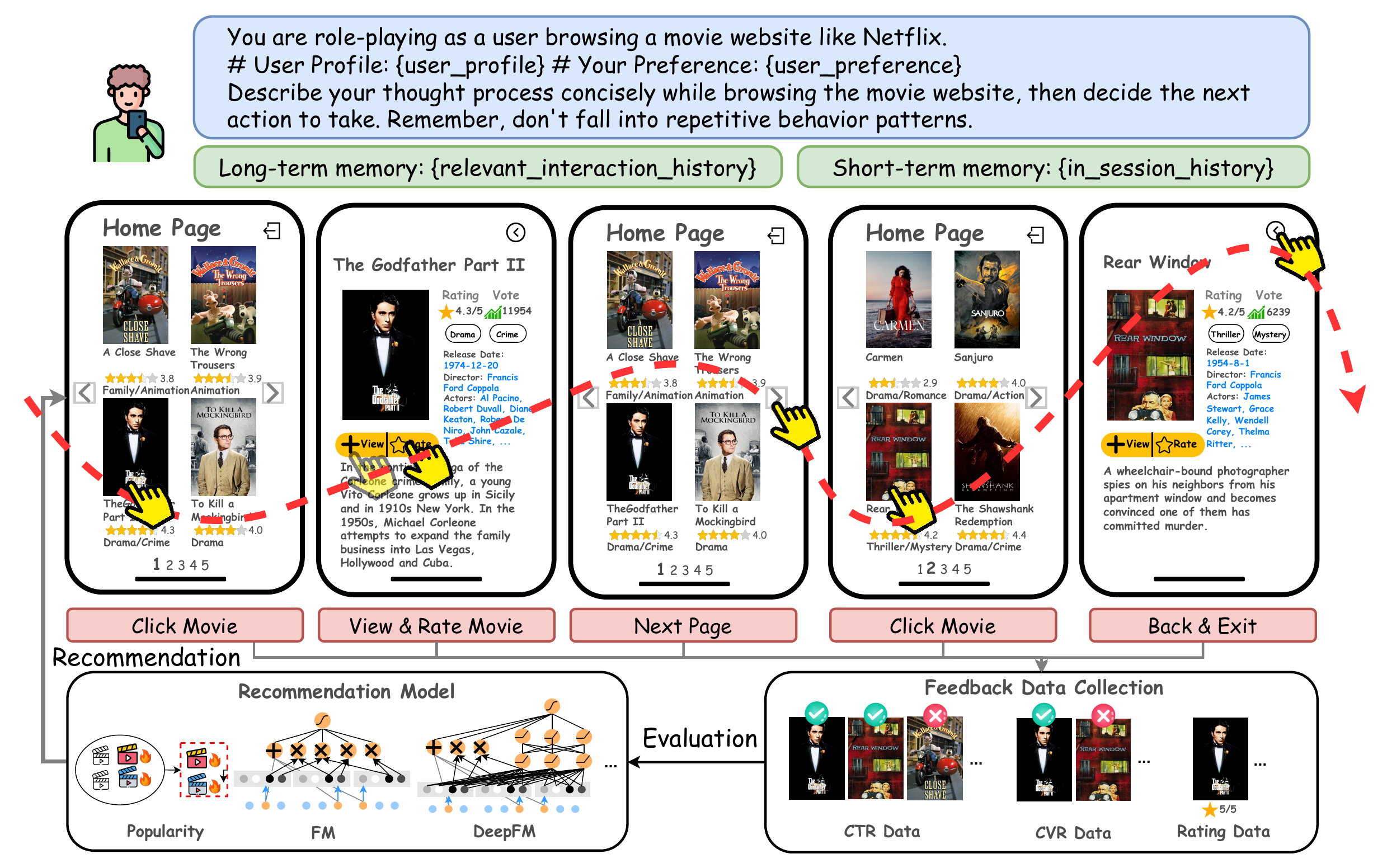}
        \caption{The recommendation sandbox environment comprises two key components: recommendation algorithms and a user interface. The recommendation algorithms generate recommendation lists displayed to users on the home page and individual movie detail pages. User interaction with this interface, including click-through rate (CTR), conversion rate (CVR), and ratings, provides data for recommendation model evaluation.}
\label{fig:Fig1_UserInterface}

\end{figure*}

To simulate a realistic movie recommendation environment, we design a multimodal interactive user interface that emulates popular platforms such as Netflix and IMDb, as illustrated in Figure~\ref{fig:Fig1_UserInterface}.

\subsubsection{\textbf{User Interface Platform}}
We design a recommendation user interface (UI) to simulate a real-world movie environment, featuring a home page and a movie detail page. Users can seamlessly navigate between these pages, each providing different levels of movie details and unique interactive options.

\noindent \textbf{Home Page.} As depicted in Figure~\ref{fig:Fig1_UserInterface}, the home page serves as the initial interface users encounter upon visiting the site. Here, users can view concise information about each movie, including the poster, title, rating, and genre. This interface enables users to efficiently browse through multiple movies and select those of interest. Available actions on this page include navigating to the next or previous page and clicking on a movie to access more detailed information.

\noindent \textbf{Movie Detail Page.} 
Upon selecting a movie of interest, users are directed to the Movie Detail Page, as shown in Figure~\ref{fig:Fig1_UserInterface}. This dedicated page offers comprehensive information about the chosen movie, significantly enriching the user's understanding through a plot overview and detailed metadata such as vote count, release date, director, and cast, in addition to the information available on the home page. A high-resolution poster is also provided. This extensive and granular information enables users to make informed decisions about whether to watch the movie or return to the home page. Users have the option to watch the movie, rate it, or navigate back to the home page.

\subsubsection{\textbf{Integration with Recommendation Algorithms}}
The environment supports the integration of various recommendation algorithms, as shown in Figure~\ref{fig:Fig1_UserInterface}, offering scalability for developing and optimizing recommender systems. It includes collaborative filtering algorithms such as random recommendation, popularity-based models~\cite{steck2011item}, Factorization Machines (FM)~\cite{rendle2010factorization}, DeepFM~\cite{guo2017deepfm}, AFN~\cite{cheng2020adaptive}. Model performance is evaluated using Click-Through Rate (CTR, the ratio of clicks to impressions on the home page), Conversion Rate (CVR, the ratio of movie detail page views), and Average Rating (AR) data collected from user simulation feedback.

\begin{figure*}
	\centering
	\includegraphics[width=1.0\linewidth]{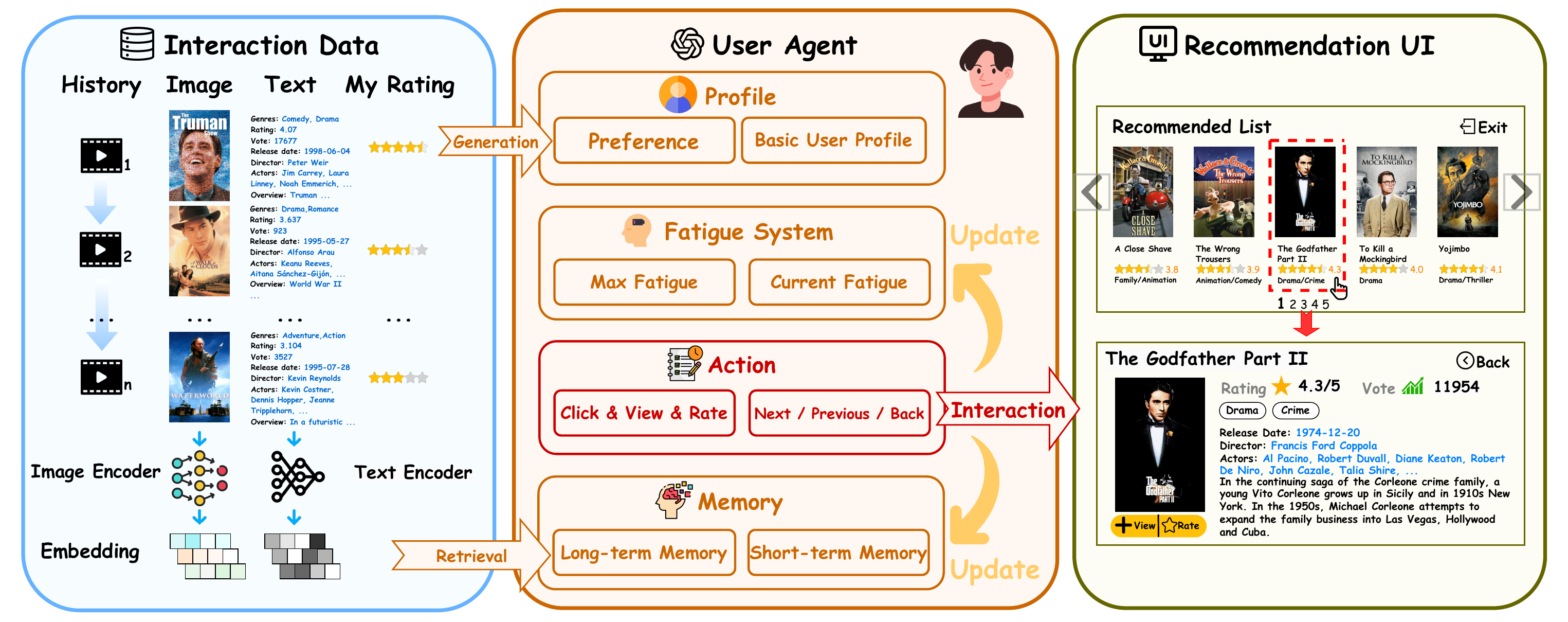}
	\caption{{The Framework of \name. Our agent design involves three components: multimodal User Agent (Orange Section), Recommendation UI (Green Section), and Interaction Data (Blue Section). The Recommendation UI provides a multimodal, multi-interface sandbox for the agent. Based on Interaction Data, the agent initializes user preferences and retrieves relevant memories. The multimodal User Agent simulates multi-page, multimodal information processing and decision-making behavior based on modules including profile, action, memory, and fatigue system.} 
    }
	\label{fig:Fig2_Agent}
\end{figure*}

\subsection{\name Architecture}
\label{section_MMAgent}
\name simulates user interaction patterns in the recommendation sandbox environment. It comprises several key components, including the agent profile module, memory module, action module, and fatigue system. The overall agent design framework is shown in Figure~\ref{fig:Fig2_Agent}.

\subsubsection{\textbf{Profile Module}}
To facilitate the agent's ability to perceive varying granularities of movie information, it is essential for the agent profile to delineate fine-grained preferences that inform the agent's core behavior patterns and decision-making logic~\cite{li2024personal}.
The profile module consists of two key elements: the user profile and user preferences. The user profile includes demographic details such as gender, occupation, age, and location. The user preferences tailored to specific movie characteristics are summarized by LLMs based on the user's interaction history through prompting. 

\subsubsection{\textbf{Memory Module}}
The memory module is essential for retaining interaction history and supporting decision-making in recommendation systems~\cite{huang2024understanding}. While existing memory designs provide a basic framework, they neglect visual modality retrieval~\cite{zhang2024generative}. To achieve the multimodal perception, we designed a memory mechanism that integrates both textual and visual retrieval. The memory module is composed of long-term and short-term memory, where long-term memory stores historical interaction records, and short-term memory captures interactions within the current session.

\noindent \textbf{Long-Term Memory.}  
To model persistent user preferences beyond the current session, we design a long-term memory module that stores and retrieves historical interaction data across sessions. This component captures stable interests over time and complements the short-term memory, which focuses on in-session context.
(1)~\textit{Textual Retrieval:} We use the \texttt{text-embedding-3-small} model~\cite{textembedding3small} to encode rich movie meta-information into textual embeddings \(\boldsymbol{e}_{\text{text}}\). When exposed to new content, the agent formulates textual queries \(\boldsymbol{q}_{\text{text}}\) based on the interface description and retrieves semantically aligned records from past sessions, reinforcing consistent long-term preferences.
(2)~\textit{Visual Retrieval:} We further incorporate visual memory by encoding movie posters using CLIP~\cite{radford2021learning} into visual embeddings \(\boldsymbol{e}_{\text{image}}\). The agent generates visual queries \(\boldsymbol{q}_{\text{image}}\) informed by elements such as color tones and object layouts, and retrieves relevant visual memories via cosine similarity.
By integrating both textual and visual retrieval mechanisms, the long-term memory module robustly supports human-like interaction recall across modalities and sessions, thereby enabling stable and consistent user modeling over extended periods.

\noindent \textbf{Short-Term Memory.}  
To support context-aware decision-making within a single session, we design a short-term memory module that continuously tracks all in-session interactions between the agent and the environment. Unlike long-term memory, which captures cross-session preferences and aggregates persistent behavioral patterns, this component focuses specifically on in-context observations and moment-to-moment behaviors to maintain coherence and avoid redundancy during exploration. Each interaction is stored in a structured format, including the interface type (e.g., homepage, movie detail page), local observations (e.g., whether certain items attract attention), estimated interest levels (on a 1–5 scale), and the corresponding action taken by the agent. This design enables the agent to reason over recent events, refer to its recent decisions, and dynamically adjust future actions accordingly, thereby better mimicking the sequential and adaptive nature of human short-term awareness and decision-making during a recommendation session.

\subsubsection{\textbf{Action Module}}
The action module defines the agent's workflow and permissible actions within the recommendation environment~\cite{zhang2024generative}, adapting to diverse interactive interfaces.  The agent's workflow involves retrieving relevant memories, analyzing the current page, and contextually determining its next action.  After each action, the agent's memory is updated, and the environment responds with page transitions and updated content.  Actions are interface-specific. For example, on the home page, the agent can \textit{click} a movie for details or navigate using \textit{next page} or \textit{previous page}. On the movie detail page, the agent can \textit{view}, \textit{rate} or \textit{back}.  

\subsubsection{\textbf{Fatigue System}}
Although the agent can achieve personalized feedback with the recommendation environment based on its profile, memory, and action modules, it often faces the issue of excessive exploration across multiple interfaces, leading to inconsistency with real user behavior. To address this discrepancy, we propose a fatigue system.

Specifically, the agent starts each session with an initial fatigue value. Each action consumes a certain amount of fatigue, which the agent considers when selecting actions. When the agent's fatigue value reaches zero, it will actively exit the recommendation environment. We categorize actions based on the frequency with which real users perform them as follows: (1) High-frequency actions, which include browsing behaviors such as previous page, next page, exit, and back; (2) Medium-frequency actions, which involve clicking to explore movies of interest; and (3) Low-frequency actions, which consist of watching and rating movies to confirm interest.
The fatigue cost is determined by two factors including the type of action and the agent's level of interest in the current page, which can be defined as 
\begin{equation}
F = C_a \cdot \left( \phi_{\text{max}} - \frac{(\iota - \iota_{\text{min}})(\phi_{\text{max}} - \phi_{\text{min}})}{\iota_{\text{max}} - \iota_{\text{min}}} \right),
\end{equation}
\noindent where:
 \( F \) is the computed fatigue cost,
 \( C_a \) represents the base fatigue coefficient for the action,
 \( \phi_{\text{max}} \) and \( \phi_{\text{min}} \) are the maximum and minimum fatigue modifiers, respectively,
 \( \iota \) denotes the current interest level,
 \( \iota_{\text{max}} \) and \( \iota_{\text{min}} \) are the maximum and minimum interest levels, respectively.

The accumulated fatigue value influences the agent's behavior. As fatigue increases, the agent becomes more likely to engage in less demanding activities or to completely exit the session. 

\begin{table*}[t]
\normalsize
    \centering
    \caption{Performance of recommendation model evaluation within the \name framework, across ML and Fashion datasets, including Gemini-2.5-flash as an additional backbone. The best result is \textbf{bolded}, and the second-best is \underline{underlined}.}
\label{tab:tab1_ABTest_Model}
    \begin{tabular}{@{}c|c|ccc|ccc|ccc|cc@{}}
\toprule
\multirow{2}{*}{Dataset} & \multirow{2}{*}{Model} 
& \multicolumn{3}{c|}{\name (GPT-4o)} 
& \multicolumn{3}{c|}{\name (GPT-4o-mini)} 
& \multicolumn{3}{c|}{\name (Gemini-2.5-flash)} 
& \multicolumn{2}{c}{Real-World} \\

                         &                        & CTR     & CVR     & AR     
                         & CTR     & CVR     & AR     
                         & CTR     & CVR     & AR     
                         & Recall  & NDCG    \\
\midrule

\multirow{5}{*}{ML} 
& Random                & 0.2330 & 0.1147 & 4.27 & 0.0872 & 0.0461 & 4.01 & 0.2905 & 0.1863 & 4.01 & 0.0066 & 0.0222 \\
& Pop                   & 0.3077 & 0.1835 & 4.30 & 0.1886 & 0.1181 & 4.20 & 0.4520 & 0.2920 & 4.12 & 0.0216 & 0.0881 \\
& FM                    & 0.3635 & 0.2940 & 4.70 & 0.2642 & 0.1984 & 4.52 & 0.5000 & 0.3855 & 4.43 & 0.0353 & 0.0987 \\
& AFN                   & \underline{0.4301} & \underline{0.3385} & \underline{4.72} & \underline{0.2845} & \underline{0.1995} & \underline{4.52} & \underline{0.5313} & \underline{0.4225} & \underline{4.52} & \underline{0.0422} & \textbf{0.1238} \\
& DeepFM                & \textbf{0.4453} & \textbf{0.3458} & \textbf{4.75} & \textbf{0.2891} & \textbf{0.2094} & \textbf{4.52} & \textbf{0.5417} & \textbf{0.4333} & \textbf{4.57} & \textbf{0.0429} & \underline{0.1130} \\
\midrule

\multirow{5}{*}{Fashion} 
& Random                & 0.0158     & 0.0073     & 4.42 & 0.0031 & 0.0008 & 4.12   & 0.0426    & 0.0093     & 4.40     & 0.0079  & 0.0028 \\
& Pop                   & 0.0561     & 0.0098     & 4.46    & 0.0212 & 0.0008 & 4.32   & 0.0937     & 0.0158     & 4.44   & 0.0157  & 0.0059 \\
& FM                    & 0.0612     & 0.0108     & 4.50    & 0.0311 & 0.0067 & 4.5  & 0.1052     &  0.0285    & 4.50     & 0.0591  & 0.0212 \\
& AFN                   & \underline{0.0754}     & \underline{0.0158}     & \underline{4.50}    & \underline{0.0407} & \underline{0.0081} & \underline{4.50} & \underline{0.1337}  & \underline{0.0341}   & \underline{4.52}     & \underline{0.0748} & \underline{0.0259} \\
& DeepFM                & \textbf{0.0772}     & \textbf{0.0165}     & \textbf{4.53} & \textbf{0.0414} & \textbf{0.0089} & \textbf{4.51} & \textbf{0.1409}  & \textbf{0.0385} & \textbf{4.60}     & \textbf{0.0787} & \textbf{0.0281} \\

\bottomrule
\end{tabular}
\end{table*}

\section{Experiment}
\label{section_experiment}
\subsection{Experimental Setting}
\textbf{Task and Evaluation Protocol.}  
We evaluate \name on personalized recommendation tasks in two domains: MovieLens and Amazon Fashion. User-item interactions are split chronologically into training, validation, and test sets (7:2:1). The training set is used to build recommendation models and initialize user profiles.
Evaluation is conducted from two perspectives. In the sandbox environment, we measure simulation performance using Click-Through Rate (CTR), Conversion Rate (CVR), and Average Rating (AR), which reflect the agent’s ability to mimic human behavior. To assess simulation fidelity, we compare simulated outcomes with real user feedback metrics (Recall@20 and NDCG@20~\cite{jarvelin2002cumulated}) computed on the held-out test set. The closer the simulation aligns with real-world performance, the higher the fidelity.

\textbf{Simulation Environment.}  
We build a sandbox environment that mimics real-world platforms (e.g., IMDB, Amazon), where each item includes multimodal attributes (title, genre, poster, and description). At each step, the agent views 5 items from a top-20 recommendation list and makes interaction decisions based on perceived relevance.

\textbf{Implementation Details.}  
User preferences are summarized using GPT-4o~\cite{islam2024gpt}, and we test three backbone LLMs: GPT-4o, GPT-4o-mini, and Gemini-2.5-Flash. The agent uses \textit{text-embedding-3-small} for memory retrieval and CLIP~\cite{radford2021learning} for visual encoding. DeepFM serves as the recommendation model, with ablations on data scale and feature sets.

\subsection{Recommendation Model Evaluation}
To evaluate how effectively \name simulates A/B testing within the recommender system \AB, we conducted \AB experiments from three perspectives: model comparison, data scale impact, and feature importance.

\subsubsection{\textbf{Model Comparison}}

Table~\ref{tab:tab1_ABTest_Model} reports simulation results across two domains (MovieLens and Amazon Fashion), using three backbone LLMs: GPT-4o, GPT-4o-mini, and Gemini-2.5-Flash. The experiments demonstrate that \name is effective in evaluating recommendation models under diverse configurations. Key observations are summarized below:
(1) \textbf{Reliable performance ranking}: \name captures consistent model performance ordering aligned with offline evaluation. Specifically, Pop outperforms Random due to the popularity-based strategy, FM outperforms both, and DeepFM and AFN achieve the best results with close performance. This indicates that \name can reflect meaningful preference differences among models.
(2) \textbf{Cross-domain generalization}: On Amazon Fashion, simulation results remain consistent with real-world trends, suggesting that \name generalizes well beyond the movie domain to diverse recommendation settings.
(3) \textbf{Backbone robustness}: All three backbone LLMs lead to comparable model ordering, confirming that \name is not sensitive to the specific choice of language model and can be flexibly deployed with various LLM infrastructures.
(4) \textbf{Multi-metric consistency}: \name maintains consistent model ranking across CTR, CVR, and average rating. This validates its ability to simulate realistic user behavior beyond simple user-item feedback, especially in recommendation sandbox environments with complex interaction flows.

\subsubsection{\textbf{Data Scale Impact}}

Table~\ref{tab:tab2_ABTest_Data} demonstrates the simulation results of DeepFM trained with three different proportions of data: 50\%, 75\%, and 100\%. This experiment evaluates whether \name can reflect performance variations arising from training data scale. The observation can be summarized as follows: (1) \textbf{Monotonic improvement}: As the training data increases, all three metrics(i.e. CTR, CVR, and average rating) consistently improve. This aligns with real-world evaluation results. (2) \textbf{Reliable differentiation}: \name successfully distinguishes the performance differences across data scales, indicating that the agent can capture fine-grained model improvements induced by larger datasets.

\subsubsection{\textbf{Feature Importance Evaluation}}
Table~\ref{tab:tab3_ABTest_Feature} presents a feature ablation study on the DeepFM model, where we assess the impact of different input feature sets on recommendation performance. Specifically, we train three variants:  
(1) \textit{User ID Only}, where user-side features (gender, age, occupation, zip) are removed;  
(2) \textit{Movie ID Only}, where the item-side genre feature is excluded;  
(3) \textit{All Features}, where all user and item features are retained.
Key observations are summarized below: (1) \textbf{Performance degradation under ablation}: Models trained with reduced feature sets exhibit consistent performance drops across all metrics (CTR, CVR, and average rating), confirming the importance of both user-side and item-side auxiliary features in model learning. (2) \textbf{User-side features are more informative}: The \textit{Movie ID Only} model performs better than the \textit{User ID Only} variant in CTR and CVR, suggesting that user-related features contribute more significantly to recommendation accuracy in this setting.
(3) \textbf{\name can detect fine-grained differences}: \name successfully reflects these performance gaps, demonstrating its ability to distinguish subtle variations in model input and validate feature effectiveness through simulation.

These results support the utility of \name as a reliable simulation framework for evaluating recommender system performance across various aspects, including model differentiation, data scale impact, and feature set importance.

\begin{table}[t]
    \centering
    \caption{Performance of DeepFM with various training data scale within the \name Framework. }
    \label{tab:tab2_ABTest_Data}
    \resizebox{\columnwidth}{!}{%
    \begin{tabular}{@{}cccc|cc@{}}
    \toprule
    Evaluation & \multicolumn{3}{c|}{\name(GPT-4o-mini)} & \multicolumn{2}{c}{Real-World} \\
    \midrule
    Training Data & CTR & CVR & AR & Recall & NDCG \\
    \midrule
    50\%    & 0.2205 & 0.1803 & 4.51 & 0.0275 & 0.0738 \\
    75\%    & \underline{0.2745} & \underline{0.1999} & \underline{4.51} & \underline{0.0330} & \underline{0.0918} \\
    100\%   & \textbf{0.2891} & \textbf{0.2094} & \textbf{4.52} & \textbf{0.0429} & \textbf{0.1130} \\
    \bottomrule
    \end{tabular}
    }
\end{table}

\begin{table}[t]
    \centering
    \caption{Performance of DeepFM with various features within the A/B Agent Framework. }
    \label{tab:tab3_ABTest_Feature}
    \resizebox{\columnwidth}{!}{%
    \begin{tabular}{cccc|cc}
    \toprule
    Evaluation & \multicolumn{3}{c|}{\name(GPT-4o-mini)} & \multicolumn{2}{c}{Real-World} \\
    \midrule
    Feature & CTR & CVR & AR & Recall & NDCG \\
    \midrule
    User ID Only  & 0.2754 & 0.1982 & 4.47 & 0.0359 & 0.0981 \\
    Movie ID Only & \underline{0.2850} & \underline{0.2097} & \underline{4.51} & \underline{0.0372} & \underline{0.0966} \\
    All           & \textbf{0.2891} & \textbf{0.2094} & \textbf{4.52} & \textbf{0.0429} & \textbf{0.1130} \\
    \bottomrule
    \end{tabular}
}
\end{table}

\subsection{Agent Alignment}
\subsubsection{\textbf{User Taste Alignment}}
\begin{figure}
    \makebox[\linewidth][c]{%
        \includegraphics[width=0.85\linewidth]{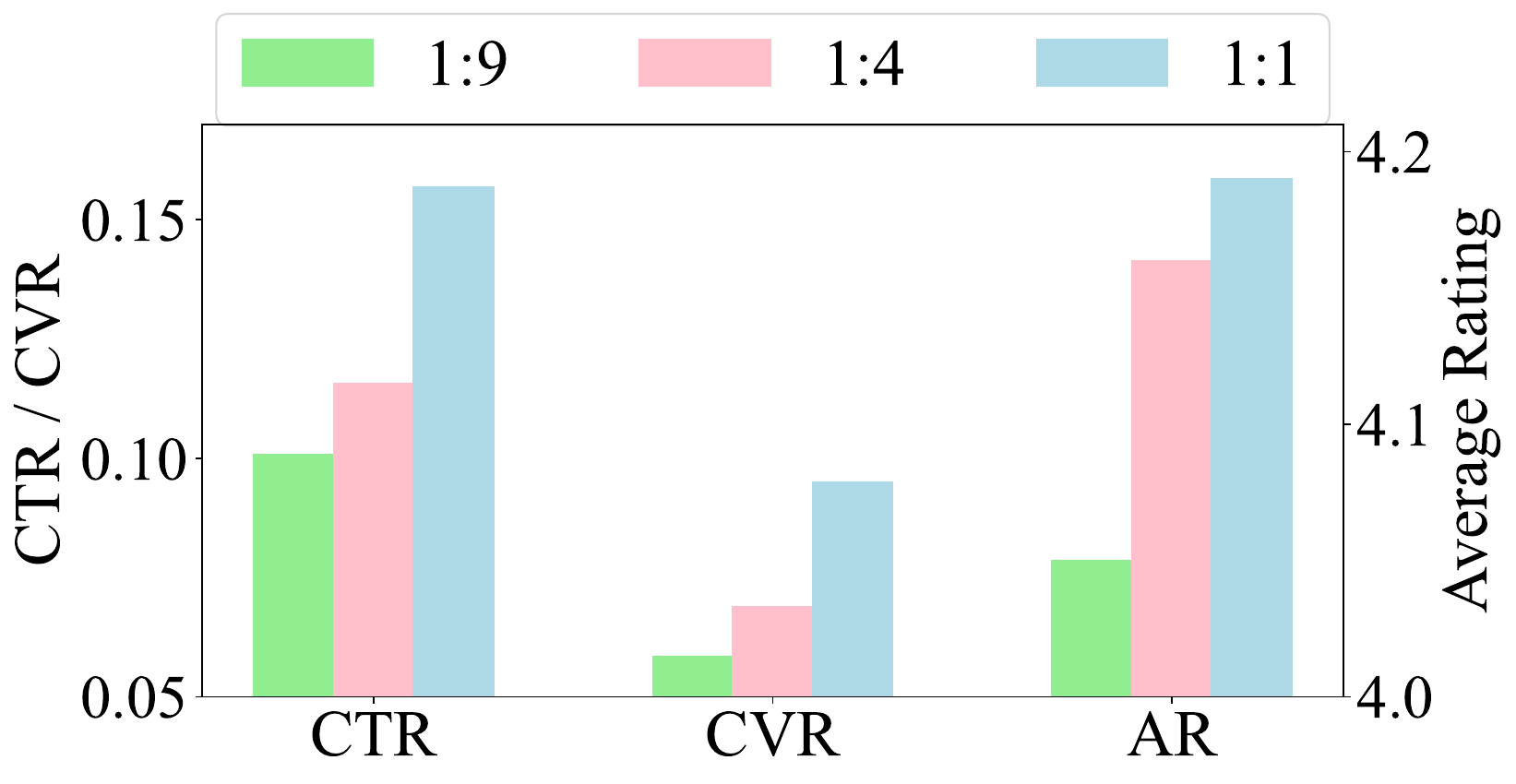}
    }
    \caption{Impact of positive recommendation ratio on CTR, CVR, and Rating.}
    \label{fig:Fig4_user_pre_alin}
\end{figure}

To validate the alignment between simulated preferences and real user behavior, we conduct a user taste alignment study. The hypothesis is that if the simulator captures real user interests, it should show stronger acceptance for recommendation lists containing a higher proportion of ground-truth positive items.
We construct three sets of recommendation lists with positive-to-negative ratios of 1:9, 1:4, and 1:1, respectively. For each user, we include 20 items composed of real clicked movies and sampled negatives. We then evaluate how \name responds to each recommendation.
As shown in Figure~\ref{fig:Fig4_user_pre_alin}, the agent exhibits consistently higher CTR, CVR, and AR under recommendation lists with more positive items, suggesting strong alignment with real user preferences.
\subsubsection{\textbf{Activity Trait Alignment}}
\begin{figure}
    \makebox[\linewidth][c]{%
        \includegraphics[width=1.0\linewidth]{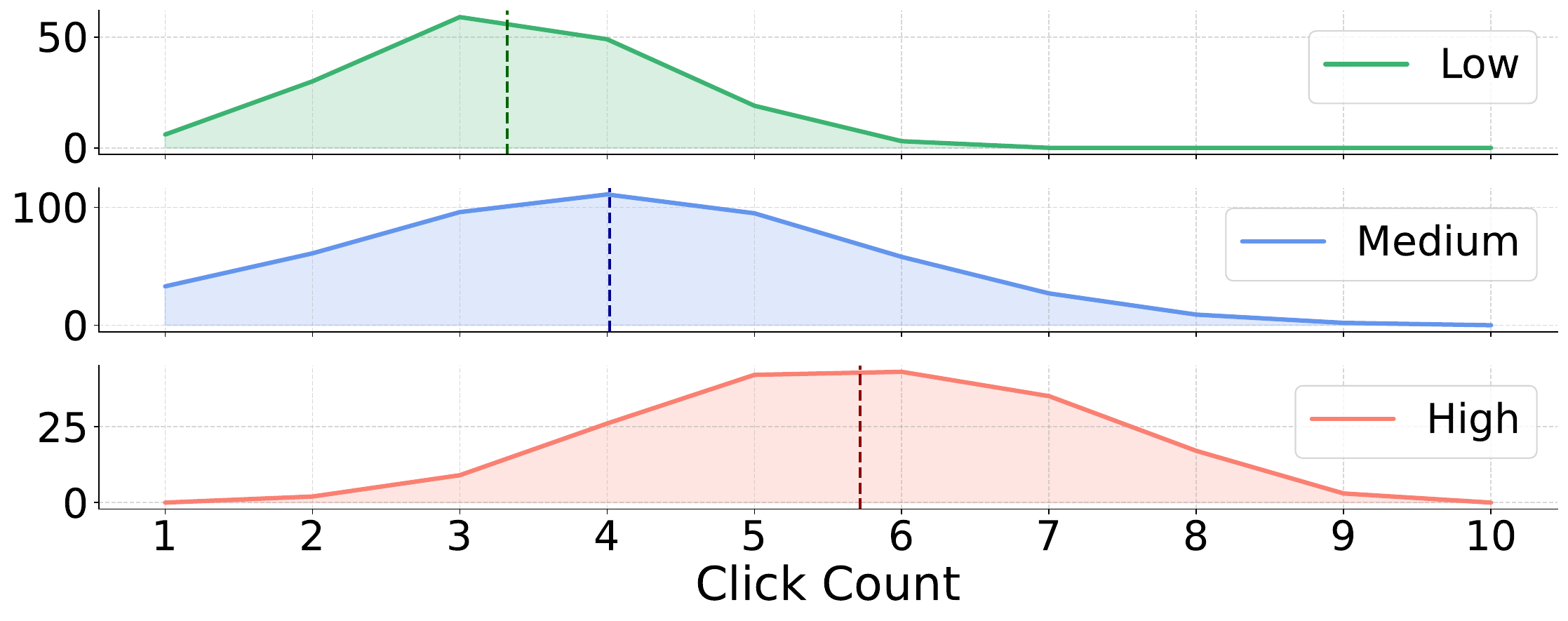}
    }
    \caption{Click distribution among user groups with different activity traits.}
    \label{fig:Fig5_clickcount}
\end{figure}

We assign each \name an activity trait (low, medium, or high) to control its overall engagement level. To verify whether the agent's behavior aligns with its assigned trait, we analyze the distribution of movie clicks under each setting.
As shown in Figure~\ref{fig:Fig5_clickcount}, agents with different activity levels exhibit distinct click distributions. All groups roughly follow a Gaussian distribution, but the mean number of clicks shifts from 3.2 (low) to 4.0 (medium) and 5.8 (high). This consistent increase in click volume confirms that the agent's behavior aligns well with its assigned activity trait, demonstrating the effectiveness of our fatigue-aware design in modulating user engagement.

\begin{table*}[]
    \centering
     \caption{The AUC comparison between various models for data augmentation experiment. It is worth noting that an AUC increase of 0.001 can be considered a significant improvement in CTR prediction~\cite{li2022inttower} The best result is \textbf{bolded}, and the second-best is \underline{underlined}.}
\label{tab:tab4_dataaugmentation}
\begin{tabular}{@{}ccccccccc@{}}
\toprule
Model & NFM & xDeepFM & Wide\&Deep & DCN & DeepFM & AFN & AutoInt & PNN \\ \midrule
Original & 0.7438 & 0.7478 & 0.7469 & 0.7517 & 0.7520 & 0.7512 & 0.7510 & 0.7474 \\
+Sim. Click Data(w/o vision) & 0.7458 & 0.7482 & 0.7485 & 0.7533 & 0.7501 & 0.7502 & 0.7527 & 0.7503 \\
+Sim. View Data(w/o vision) & 0.7455 & 0.7486 & 0.7482 & 0.7534 & 0.7528 & \underline{0.7543} & \underline{0.7532} & \underline{0.7518} \\ \midrule
+Sim. Click Data(w/ vision) & \underline{0.7464} & \underline{0.7502} & \underline{0.7501} & \textbf{0.7539} & \underline{0.7541} & 0.7530 & 0.7531 & 0.7499 \\
+Sim. View Data (w/ vision) & \textbf{0.7475} & \textbf{0.7517} & \textbf{0.7521} & \underline{0.7537} & \textbf{0.7542} & \textbf{0.7556} & \textbf{0.7543} & \textbf{0.7520} \\ 
\bottomrule
\end{tabular}

\end{table*}




\subsection{Data Augmentation}
To assess the effectiveness of \name-generated interaction data, we simulate agent behaviors under DeepFM-based recommendations and collect two types of signals: 2,518 homepage clicks and 1,884 viewing records from movie detail pages. Although these simulated samples (~4K) are significantly smaller than the original training set (700K), they are integrated with the original dataset to train downstream recommenders.
Table~\ref{tab:tab4_dataaugmentation} presents the offline AUC performance across eight representative models, including NFM~\cite{he2017neural}, xDeepFM~\cite{guo2017deepfm}, Wide\&Deep~\cite{cheng2016wide}, DCN~\cite{wang2017deep}, DeepFM~\cite{guo2017deepfm}, AFN~\cite{cheng2020adaptive}, AutoInt~\cite{song2019autoint}, and PNN~\cite{qu2016product}.
We observe consistent performance gains across all models, especially when using simulated data from vision-aware \name. On click data, AUC improves by more than 0.002 across most models, with Wide\&Deep achieving the highest gain of 0.0032. On view data, the improvements are even more pronounced: NFM, xDeepFM, and DeepFM achieve gains of 0.0037, 0.0039, and 0.0022, respectively. AFN, AutoInt, and PNN all show their best results with vision-aware view data.
These results demonstrate that even a small amount of high-quality simulation data can lead to measurable improvements in model performance, highlighting the utility of \name in enhancing recommendation systems through efficient data augmentation.

\subsection{Ablation Study}
To evaluate the contribution of visual modality in agent behavior simulation, we conduct an ablation study by removing poster images from the UI interface during simulation. Under this setting, agents receive the same recommendation list (from DeepFM) but without visual exposure to movie posters. The resulting interaction data, denoted as “(w/o vision)” in Table~\ref{tab:tab4_dataaugmentation}, is used for training downstream recommenders.
As shown in the table, while vision-agnostic simulation still leads to marginal improvements over the original training set, the gains are consistently smaller compared to vision-aware settings. For example, with simulated click data, DeepFM improves from 0.7520 to 0.7541 with visual input, while only reaching 0.7501 without it. Similar trends hold across other models and interaction types.
These results confirm that visual modality plays a critical role in producing realistic user-agent behavior. The agent's interaction patterns are better aligned with user preferences when visual information is available, leading to more effective data augmentation.

\subsection{Case Study}
\begin{table*}[t]
\centering
\caption{Case analysis of \name simulation: green highlights preference/acceptance-related info; red highlights rejection/disinterest cues
}
\label{tab:case_analysis}
\small
\renewcommand{\arraystretch}{1.25}
\begin{tabular}{@{} c p{3.5cm} p{3.5cm} p{3.5cm} p{3.5cm} @{}}
\toprule
\raisebox{2.02cm}{\textbf{Target Items}}
& \includegraphics[width=3.8cm]{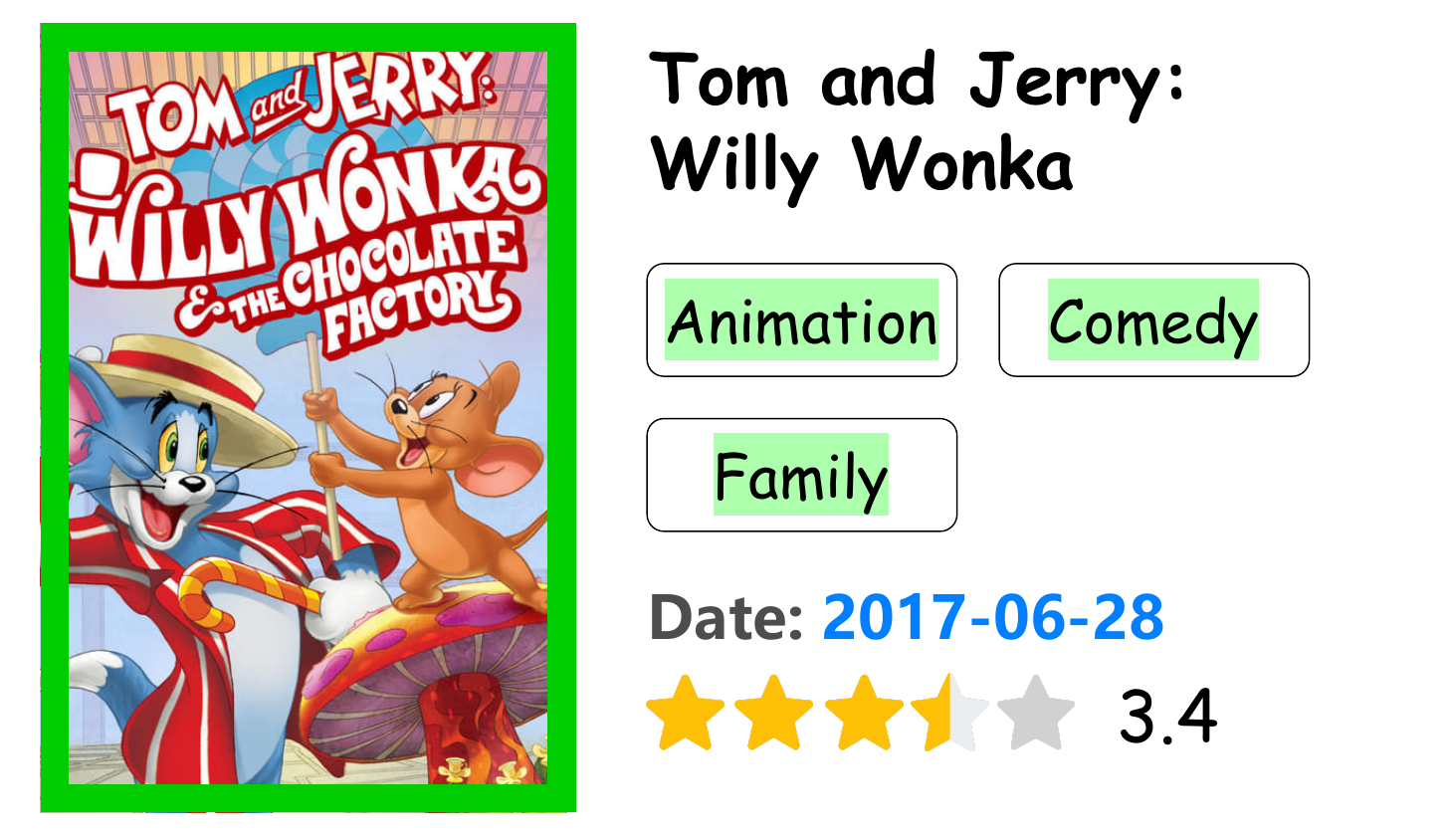} 
& \includegraphics[width=3.8cm]{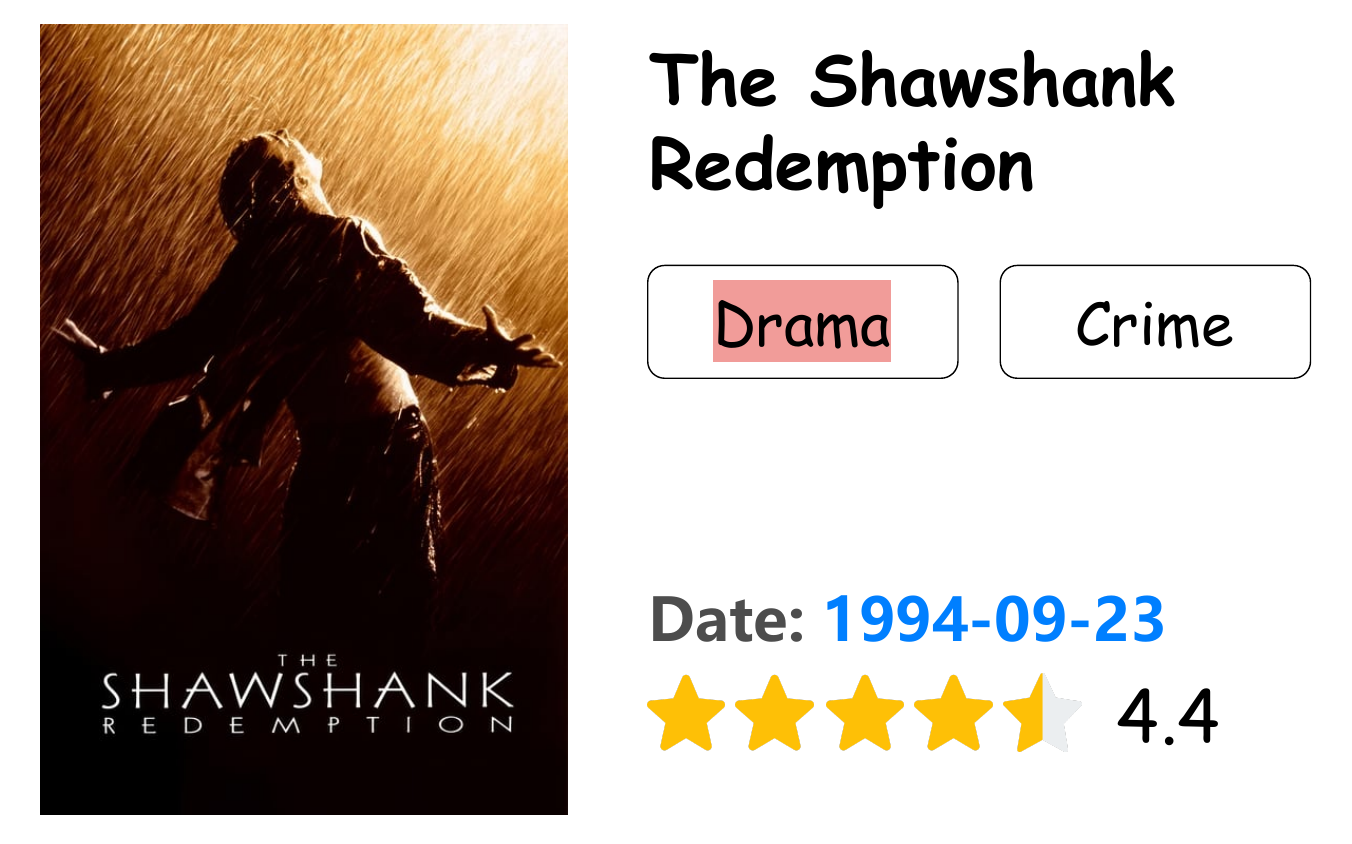} 
& \includegraphics[width=3.8cm]{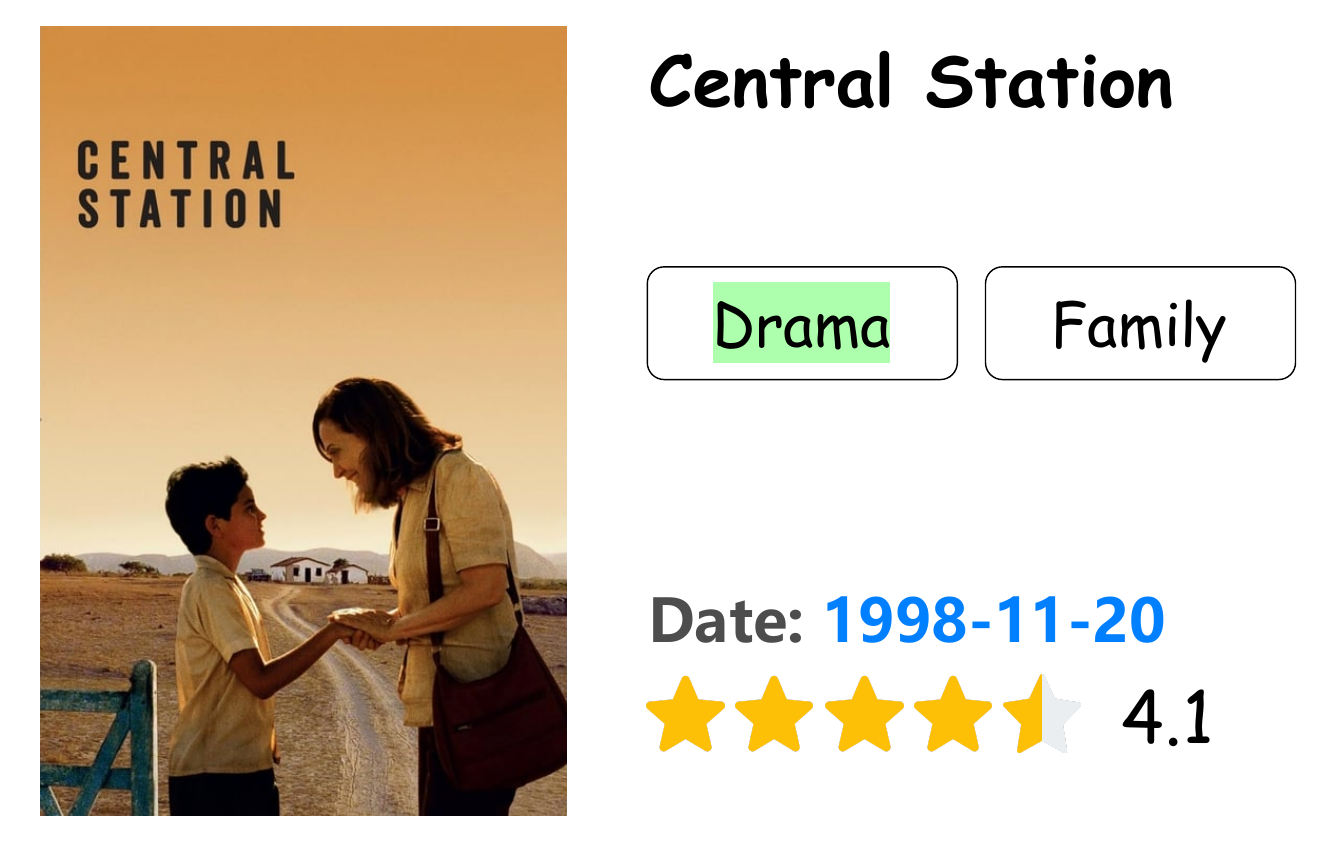} 
& \includegraphics[width=3.8cm]{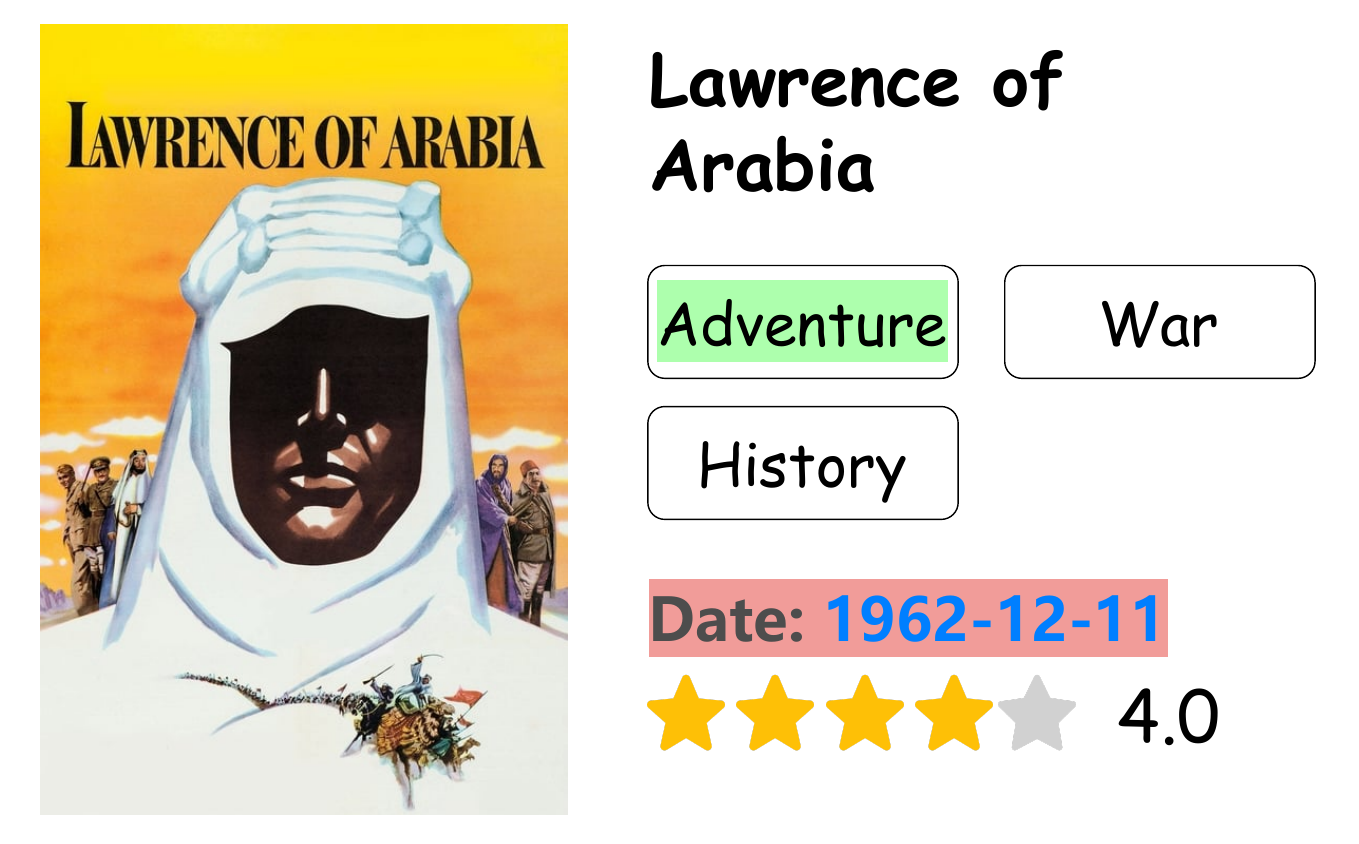}
\\
\midrule
\textbf{Action} 
& Click 
& Next Page 
& Watch \& Rate 5.0
& Exit \\
\midrule
\textbf{Profile} 
& Prefers Comedy, Family, Animation. Colorful posters
& Limited interest in drama 
& Prefers Drama and Romance (5 stars);
& Prefers Adventure films from the 80s and 90s \\
\midrule
\textbf{Memory} 
& Rated \textit{The Addams Family} 4.0 \newline
Rated \textit{Last Action Hero} 4.0
& Rated \textit{Solo} 2.0 \newline
Rated \textit{King Kong} 3.0
& Rated \textit{Breaking the Waves} 5.0 \newline Rated \textit{Three Colors: Red} 5.0
& Click \textit{The Third Man} \newline \colorbox{red!15}{Page 3 don’t fit my tastes}.. \\
\midrule
\textbf{Fatigue} 
& 0.0/30 $\rightarrow$ 2.0/30 
& 0.0/30 $\rightarrow$ 2.0/30
& 9.5/30 $\rightarrow$ 19.5/30 
& \colorbox{red!15}{25.4/30} $\rightarrow$ 30.0/30 \\
\midrule
\textbf{Explanation} 
& \colorbox{green!15}{Like poster and genres} 
& \colorbox{red!15}{Drama not my taste} 
& \colorbox{green!15}{Matches past high ratings} 
& \colorbox{red!15}{Old and fatigue, not interested} \\
\bottomrule
\end{tabular}
\end{table*}

To qualitatively evaluate the effectiveness of \name in simulating user behavior under a multi-modal, multi-interface recommendation sandbox, case analysis for agent simulation are illustrated in Table~\ref{tab:case_analysis}. These cases illustrate key actions (click, next page, watch \& rate, exit), showing how agent behavior aligns with preferences, historical patterns, and fatigue.
Case 1 demonstrates a click decision where the agent, preferring comedy, family, and animation, is attracted to the recommended item based on its genre and vibrant poster. The agent’s action aligns with its multi-modal preferences.
Case 2 shows the agent skipping a page. Despite the high rating of The Shawshank Redemption, the agent dislikes drama and this is supported by low historical ratings for similar films. The decision is driven by genre disinterest.
Case 3 presents the agent watching and rating a movie 5.0. The agent’s preference for drama and romance is aligned with the recommendation, and past experiences reinforce the high rating.
Case 4 demonstrates an exit behavior due to a mismatch in preferences and fatigue. Although the recommended movie matches the agent’s adventure preference, its release year (1962) diverges from the agent’s preference for 80s-90s films. High fatigue (25.4/30) also leads the agent to exit.
\noindent These cases confirm that \name effectively simulates nuanced user behaviors, accounting for preferences, rejection cues, and fatigue-induced disengagement.

\subsection{Comparison with Existing Simulators}

\begin{table}[t]
  \centering
  \caption{Comparison of user simulators}
  \label{tab:simulator_comparison}
  \resizebox{\columnwidth}{!}{
      \begin{tabular}{lccc}
        \toprule
        \textbf{Simulator} & \textbf{Simulation} & \textbf{Multi-modal} & \textbf{Evaluation} \\
        \midrule
        RecoGym~\cite{rohde2018recogym} & Direct-Response & \ding{55} & Case Study \\
        RecSim~\cite{ie2019recsim} & Direct-Response  & \ding{55} & Case Study \\
        VirtualTaobao~\cite{shi2019virtual}    & Simplified UI      & \ding{55}     & Online  \\
        Adversarial~\cite{chen2019generative}  & Simplified UI         & \ding{55}           & Offline  \\
        KuaiSim~\cite{zhao2023kuaisim} & Simplified UI             & \ding{55}       & Offline  \\
        SUBER~\cite{corecco2024suber} & Simplified UI & \ding{55}                  & Offline  \& Case Study \\
        Agent4Rec~\cite{zhang2024generative}   & Simplified UI & \ding{55}         & Offline \& Case Study \\
        LLM-SimuRec~\cite{zhang2025llm}        & Direct-Response & \ding{55}               & Offline  \& Case Study \\
        \midrule
        \name & Sandbox UI & \ding{52}   & Offline \& Case Study \\
        \bottomrule
      \end{tabular}
    }
\end{table}

Table~\ref{tab:simulator_comparison} summarizes representative user simulators across simulation interface, modality, and evaluation.
Early works like RecoGym~\cite{rohde2018recogym} and RecSim~\cite{ie2019recsim} adopt simple direct-response setups with limited interaction fidelity. Later simulators such as VirtualTaobao~\cite{shi2019virtual}, Adversarial~\cite{chen2019generative},  KuaiSim~\cite{zhao2023kuaisim}, and Agent4Rec~\cite{zhang2024generative} introduce richer interaction flows, but remain text-only and overlook visual signals prevalent in real-world platforms.
LLM-SimuRec~\cite{zhang2025llm} leverages LLMs for text-based simulation but lacks UI realism and multi-modal context.
In contrast, \name integrates a sandbox-style UI with full multi-modal support, including posters and metadata. It enables more realistic and explainable user modeling, and is evaluated using both offline metrics and case-based analysis.
By bridging the gap between interaction fidelity and modality grounding, \name sets a new standard for user simulation in recommendation research.

\section{Related Work}

\subsection{Traditional User Simulator}
Traditional user simulators set the behavior mode based on rules, or use GAN and reinforcement learning to model user behavior~\cite{ie2019recsim, rohde2018recogym, shi2019virtual, chen2019generative, bai2019model}. RecSim~\cite{ie2019recsim} configures the simulation environment based on rules, including user preferences, user status, item similarity, and recommendation models, \etc to simulate the sequential interaction between users and items. UserSim~\cite{zhao2021usersim} uses GAN to train user simulators, uses generators to capture the distribution of user historical log behavior, and uses discriminators to distinguish between real and fake user logs. 
Traditional user simulators have two main drawbacks: 1) They rely on predefined rules or require large amounts of data to train user simulators, which can be resource-intensive and less adaptable; 2) They often lack interpretability, making it difficult to understand the reasoning behind simulated user behaviors.


\subsection{LLM User Simulator}
Equipped with prior open-world knowledge, LLM-based user simulators can provide flexible interaction feedback and explainable thought processes~\cite{xi2023rise}. Several works apply LLM-based user simulators in recommender systems. ToolRec~\cite{zhao2024let} designs a user simulator to evaluate user preference and provide recommendations by tool learning. RecAgent~\cite{wang2023user} and S$^{3}$~\cite{gao2023s} introduce a recommendation agent with social networks environment. Agent4Rec~\cite{zhang2024generative} develops a user simulator to interact with movie websites in a page-by-page manner. In addition, some works also utilize the LLM-based user simulator to evaluate the conversational recommender system~\cite{wang2023rethinking, Yang_2024, yoon2024evaluating, zhu2024reliable}.
However, these works lack the ability to use image-modal information and interact effectively with a realistic recommendation environment, limiting the agent's capability to simulate real-world user behavior.

\section{Conclusion}
In this paper, we propose a novel multimodal LLM-based user agent framework \name for \AB. Recognizing the limitations of current agents in replicating human perception and interaction within realistic environments, we first constructed a comprehensive multimodal and multi-page recommendation sandbox that mirrors online platform UIs. Within this environment, \name leverages multimodal information perception, fine-grained user preferences, and integrates profiles, memory, and a fatigue system to simulate complex human decision-making. 
Through extensive recommendation system evaluation and data augmentation experiments, we demonstrate the strong potential of \name to simulate user behavior in \AB sandbox environment.

\section*{Acknowledgements}
This research was partially supported by National Natural Science Foundation of China (No.62502404), Hong Kong Research Grants Council (Research Impact Fund No.R1015-23, Collaborative Research Fund No.C1043-24GF, General Research Fund No.11218325), Institute of Digital Medicine of City University of Hong Kong (No.9229503), and Huawei (Huawei Innovation Research Program).

\section*{Statement on Copyrights}
The movie poster images displayed in this paper are utilized strictly for academic research and illustrative purposes to demonstrate the recommendation scenarios. We acknowledge that the copyright of these visual assets belongs to their respective production studios or distributors, and their inclusion here falls under the fair use principles for academic publication.


\section*{Discussion}
Despite the potential of \name, we acknowledge limitations regarding simulation scope and model constraints. First, our framework operates within a closed environment, overlooking external real-world influences like social media and peer interactions that shape user decisions. Second, intrinsic LLM issues, such as hallucinations and repetitive behaviors, may compromise simulation reliability. Future work will focus on integrating broader information channels and mitigating these generative anomalies. 
\bibliographystyle{ACM-Reference-Format}
\balance
\bibliography{9Reference}

\clearpage
\appendix
\section{More Implementation Details of \AB}
\label{sec:implement-detail}

In this section, we firstly detailed the \textbf{fatigue setting} on \name in Table~\ref{tab:fatigue_values}. 

\begin{table}[H] 
    \centering
    \caption{Fatigue values for different actions in GPT-4o and GPT-4o-mini}
    \label{tab:fatigue_values}
    \begin{tabular}{lccc}
    \toprule
    \multirow{2}{*}{Action} & \multicolumn{2}{c}{Fatigue Value} \\
    \cmidrule(lr){2-3}
     & GPT-4o & GPT-4o-mini \\
    \midrule
    click\_movie & 15 & 2 \\
    watch\_and\_rate\_movie & 40 & 10 \\
    previous\_page & 2 & 2 \\
    next\_page & 2 & 2 \\
    back\_action & 2 & 5 \\
    exit\_action & 0 & 0 \\
    \bottomrule
    \end{tabular}
\end{table}

Then, we display our prompt design in our agent framework and give a typical example of user preference construction. 

\subsection{User Preference Generation}
We provide below the prompt used for user preference generation (Figure~\ref{fig:user_pre_prompt}).

\begin{figure}[H]
    \centering
    \begin{tcolorbox}[
        colback=gray!5,
        colframe=gray!60!black,
        width=\linewidth,
        arc=2mm,
        boxrule=0.8pt,
        title=\textbf{User Preference Generation Prompt},
        fonttitle=\bfseries,
        fontupper=\small, 
        left=2pt, right=2pt, top=2pt, bottom=2pt, 
    ]
        \textbf{System Instruction:}
        You are given a user's movie interaction history. Assume the role of user. Your task is to write a clean, concise, and well-structured user preferences (taste) summary in the first person.

        \textbf{Guidelines:}
        \begin{enumerate}[nosep, leftmargin=1.2em] 
            \item \textbf{Genres:} Identify and list the genres you prefer.
            \item \textbf{Directors:} Mention directors whose works you consistently enjoy.
            \item \textbf{Actors:} Highlight actors whose performances you appreciate.
            \item \textbf{Release Date Patterns:} Note any trends in the release years of the movies you watch.
            \item \textbf{Rating Tendencies:} Describe my typical ratings for different types of movies or score ranges.
            \item \textbf{Poster Style Preference:} Based on the movie interaction history, summarize the user's preference for movie poster aesthetics, such as color schemes, compositions, and character depictions.
            \item \textbf{Conciseness:} Summarize the preferences in a manner that is clear and to the point.
        \end{enumerate}
        If no clear preferences are found, indicate this with 'not found'. Do not give any information that is not related to user preference.

        \vspace{0.2em} \hrule \vspace{0.2em}

        \textbf{Input (User History):} \texttt{\{history\}} \\
        \textbf{Response Requirement:} Remember, don't blindly repeat the contexts verbatim. \\
        \textbf{LLM Output:} \texttt{\{Summarized Preference\}}
    \end{tcolorbox}
    \caption{The prompt template for user preference generation.}
    \label{fig:user_pre_prompt}
\end{figure}

The resulting user preferences are shown in Figure~\ref{fig:user_pre_case}. The generated preferences capture fine-grained user tendencies specific to individual interaction interfaces.

\begin{figure}[H]
    \centering
    \begin{tcolorbox}[
        colback=gray!5, colframe=gray!60!black, width=\linewidth, arc=2mm, boxrule=0.8pt,
        title=\textbf{User Preference Case Study},
        fonttitle=\bfseries, fontupper=\small,
        left=2pt, right=2pt, top=2pt, bottom=2pt,
    ]
        \textbf{My Movie Preferences Summary}

        \textbf{Genres:} I have a strong preference for Drama and Romance, often enjoying films that blend these genres with elements of History, Mystery, and Crime. I also appreciate well-crafted Comedies and Thrillers.

        \textbf{Directors:} I consistently enjoy the works of Krzysztof Kieślowski, Steven Spielberg, and James Ivory. Their storytelling and direction resonate with me deeply.

        \textbf{Actors:} I appreciate the performances of Audrey Hepburn, Daniel Day-Lewis, and Helena Bonham Carter. Their ability to bring characters to life adds a significant value to the films they are in.

        \textbf{Release Date Patterns:} I tend to favor films from the 1990s, with a noticeable appreciation for classics from the 1950s to the 1970s as well.

        \textbf{Rating Tendencies:} I generally rate Drama and Romance films highly, often giving them 4 or 5 stars. Comedies and Thrillers also receive favorable ratings, typically around 4 stars. However, I am more critical of films that fail to engage me.

        \textbf{Poster Style Preference:} I prefer movie posters with classic and elegant designs, often featuring well-composed images of the main characters.

        \vspace{0.2em} \hrule \vspace{0.2em}
        
        \textbf{Summary:} In summary, my taste in movies leans heavily towards well-crafted dramas and romances, with a particular appreciation for strong performances and thoughtful direction.
    \end{tcolorbox}
    \caption{An example of a generated user preference summary.}
    \label{fig:user_pre_case}
\end{figure}

\section{Recommendation System Implementation}

We implement all models using the DeepCTR~\cite{shen2017deepctr} framework with binary cross-entropy loss, Adam optimizer ($lr=1 \times 10^{-3}$), batch size 1024, and early stopping based on validation AUC. Unless specified otherwise, models use an embedding dimension of 64, DNN layers of (256, 128), dropout of 0.3, and L2 regularization of 0.003. 
Specific adjustments are as follows: 
FM uses no DNN with linear L2 regularization $10^{-5}$. 
DeepFM adds a third layer (64) with dropout 0.5. 
Wide \& Deep uses layers (256, 64) with dropout 0.5. 
xDeepFM and DCN (2 cross layers) use dropout 0.5 and linear L2 $10^{-5}$. 
AFN employs a log layer (256) without dropout. 
AutoInt uses two self-attention layers (2 heads). 
PNN and NFM follow the default configuration.
All models are trained on an NVIDIA GeForce RTX 4060 and converge within 20 epochs.

\end{document}